\documentclass[a4paper, oneside, twocolumn, notitlepage, 10pt]{extarticle_ecoc}
\usepackage{ecoc}
\usepackage{tikz,pgfplots}
\usepackage{subcaption}

\usetikzlibrary{fit}
\usetikzlibrary{shapes.multipart}
\usetikzlibrary{positioning}
\usepackage{caption,subcaption} % subfigures
\usetikzlibrary{plotmarks,matrix,chains,scopes,fit,calc,shapes,positioning,decorations,intersections,arrows,backgrounds,shadows}

\usetikzlibrary{external}
\usepackage{balance}
\pgfdeclarelayer{background}
\pgfsetlayers{background,main} 
\usepackage{algorithmic}
\definecolor{myDarkGreen}{rgb}{0.00000,0.58824,0.00000}%

\pgfplotsset{compat=newest}

\usepackage{stfloats}

\begin{document}
%\selectlanguage{american}    % Standard Language

\title{Increasing Achievable Information Rates via Geometric Shaping}% in Fiber Optical Communications}%

%------------------------------------------------- Authors-----------------------------------------------------%

\author{
    Bin Chen\textsuperscript{(1)}, Chigo Okonkwo\textsuperscript{(1)}, Hartmut Hafermann\textsuperscript{(2)}, Alex Alvarado\textsuperscript{(1)}
}

\maketitle                  % Create title and author

%------------------------------------------ Description of Authors ----------------------------------------------%

\begin{strip}
 \begin{author_descr}

   \textsuperscript{(1)} Department of Electrical Engineering, Eindhoven University of Technology, \href{mailto:b.c.chen@tue.nl}{\color{blue}{\underline{b.c.chen@tue.nl}}}

   \textsuperscript{(2)} Mathematical and Algorithmic Sciences Lab, Paris Research Center, Huawei Technologies France SASU, 92100 Boulogne-Billancourt, France.

 \end{author_descr}
\end{strip}

\setstretch{1.1}

%-------------------------------------------------- Abstract ---------------------------------------------------------%

\begin{strip}
  \begin{ecoc_abstract}
    % NOTE: Don't use a blank line here but start abstract right away to avoid an extra line break
Achievable information rates  are used as a metric to design novel modulation formats via geometric shaping. % and maximize throughput performance. 
The proposed geometrically shaped  256-ary constellation  achieves SNR gains of up to 1.18 dB.
%an increase in AIR of up to 7\% compared to regular square QAM.    
\end{ecoc_abstract}
\end{strip}

%-------------------------------------------------- Section 1 -------------------------------------------------------%

\section{Introduction}

% AIRs
Achievable information rates (AIRs) such as   mutual information (MI) and generalized mutual information (GMI) have emerged as practical tools to design fiber optical communication systems. AIRs have also been used to design modulations formats and to predict the performance of forward error correction (FEC)\cite{Alvarado2018_AIR}. 

A key element for maximizing the throughput in both point-to-point links as well as in flexible optical networks is the use of adaptive modulation formats and FEC. Both MI and GMI can be used to analyze this from a theoretical point of view, where GMI is usually preferred due to its practical relevance. The main drawback of the GMI is that the suboptimality of the underlying  bit-interleaved coded modulation (BICM) receiver causes a rate loss with respect to the MI. This loss can be eliminated by using BICM with iterative decoding or nonbinary FEC.

% Shaping
Signal shaping can be used in fiber optics to  close the gap to the channel capacity, either via probabilistically-shaped (PS) or geometrically-shaped (GS) constellations. In the former, the  probabilities of the constellation points are changed \cite{Buchali2016,FehenbergerJLT2016,Bocherer2017_ECOC,Buchali2017,Maher2017}, while in the latter, non-equidistant constellation points are used \cite{Qu2017,ZhangECOC2017,Kojima2017_JLT,Millar2018_OFC}. For the AWGN channel, both schemes achieve Shannon's channel capacity when the number of constellation points tends to infinity. Although PS constellations have  superior AIR performance for a finite number of constellation points with respect to GS ones \cite{BICM_book}, this technique requires the use of sophisticated coding. On the other hand, GS  rely only on properly placing the constellation points and straightforward modifications of the demapper. 

% Previous works and contributions
Recently, GS-based constellations such as GS-16QAM \cite{Qu2017}, GS-32QAM  \cite{ZhangECOC2017} and GS-APSK \cite{Kojima2017_JLT,Millar2018_OFC} have been investigated in the optical communications literature. 
GS and PS  constellations have   also been investigated in the context of the ATSC 3.0 standard\cite{Steiner2017}.
%A selection of GS constellations in  ATSC 3.0 standards have   also been compared with PS \cite{Steiner2017}.
GS constellations have even been reported to outperform PS ones in both numerical simulations \cite{Millar2018_OFC} and experiments \cite{Qu2017}. 
This is particularly noticeable  for dispersion managed links\cite{Kojima2017_JLT,Millar2018_OFC}. 
Because of this and the versatility of AIR as design performance metric, in this paper we focus on GS constellations based on both GMI and MI.
This paper systematically studies and presents GS constellations with up to $M=256$ constellation points. 
%To the best of our knowledge, this paper is the first systematic study of GS constellations with up to $M=256$ constellation points based on both MI and GMI. 
We numerically prove that these $M$-ary constellations with $M=16, 64, 256$ offer rate gains of up to $7\%$ (or equivalently, SNR gains of up to 1.18 dB) for the AWGN channel.

%-------------------------------------------------- Section 2 -------------------------------------------------------%
\vspace{-0.68em}
\section{AIR-based Geometric Shaping Optimization}
The largest AIR for a memoryless channel with complex input $X$ and output $Y$ is given by the MI, defined as %(MI) 
$I=%I(X;Y)= 
\mathbb{E}\left[\log_2{f_{Y|X}(Y|X)}/{f_Y(Y)}\right]$,
where $f_{Y|X}$ is the channel law. An AIR for BICM is the GMI, which is  defined as
$G%= \sum_{i=1}^m I(B_i;Y) 
= \sum_{i=1}^m\mathbb{E}\left[\log_2{f_{Y|B_i}(Y|B_i)}/{f_Y(Y)}\right]$, where $B_1$,$B_2$,...,$B_m$ represent the bits that are mapped to the channel input $X$, and $m=\log_2 M$ is number of bits per constellation points.

Both the MI and GMI depend on the channel SNR. The MI  depends on the location of the constellation points, and thus, the MI-based optimization problem corresponds to finding the location of $M$ points, that satisfy a power constraint (defining the SNR), and that maximize the MI. 
The GMI 
%does not only depend on the location of the points and the SNR, but also
depends also 
on the binary labeling. The GMI-based optimization problem is thus more complex than the MI one as the binary labeling also needs to be taken into account.
\begin{figure*}[h]
\centering
\includegraphics[width=1.01\linewidth]{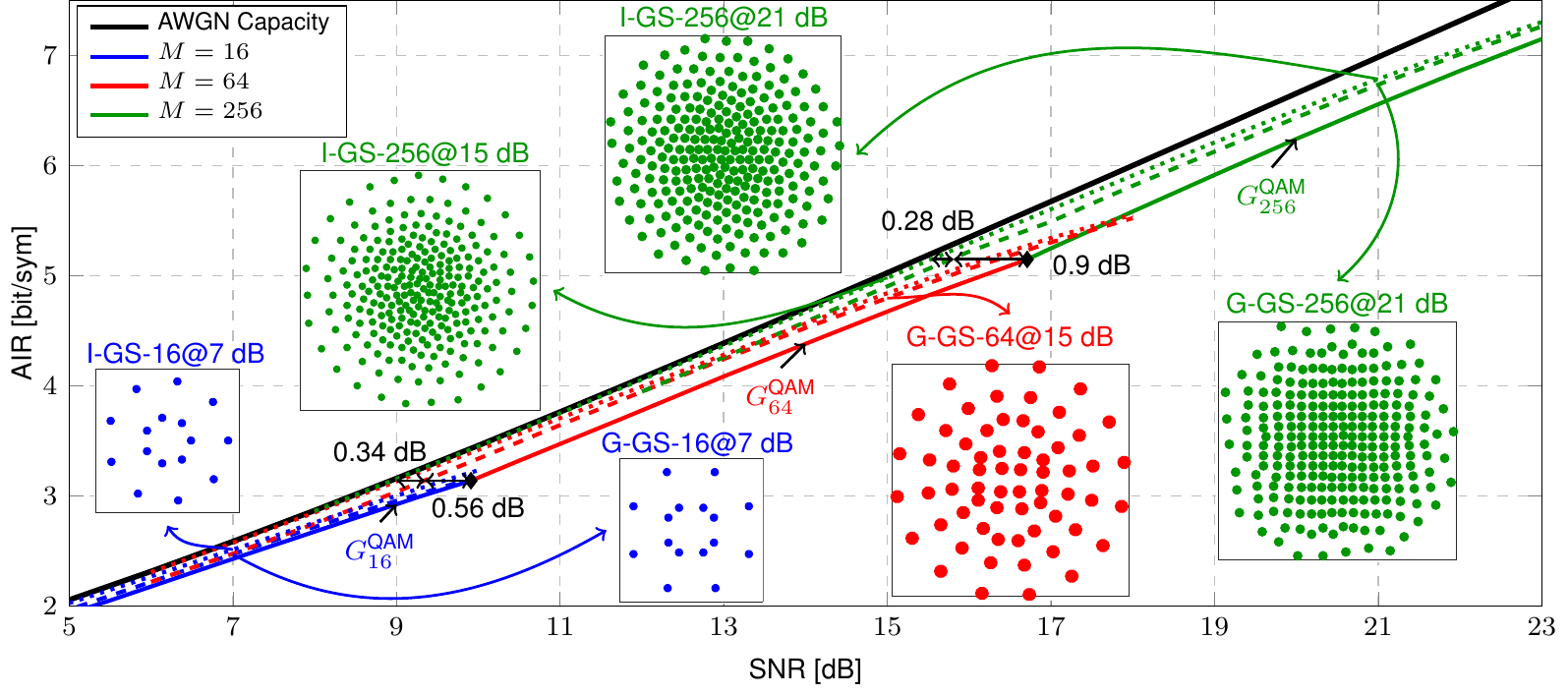}
\caption{
AIRs of MI-optimized (dotted lines) and GMI-optimized (dashed lines) constellations.
The baseline GMIs of MQAM $G_{M}^{\text{QAM}}$ as well as the AWGN capacity (black line) are also shown. Insets: GS-optimized constellations for SNR=$7,15,21$ dB.}     
\label{AIRs}
\end{figure*}
In this paper, we use a pairwise optimization algorithm\cite{Moore2009,ZhangECOC2017} to optimize the MI. For the GMI, we combine it with the binary switching algorithm \cite{Schreckenbach2003} in an iterative fashion. In both cases the optimization is repeated until the algorithm has converged or a maximum number of iterations is reached.

%\vspace{-0.65em}
\section{Numerical Results: SNR Gains}

Here we quantify the SNR gains offered by GS 

\noindent constellations with $M=16, 64, 256$ over the AWGN channel. As a baseline we use the GMI of square QAM constellations, which are known to cross each other at certain SNR values\cite{Alvarado2018_AIR,Millar2018_OFC}. This means that depending on the SNR, different modulation formats should be used. This is shown in Fig.~\ref{AIRs} (solid lines), where the black diamonds represent the switching points. These three solid lines will be used as baseline, which we denote by $G^\text{QAM}$. As shown in Fig.~\ref{AIRs}, we denote each ``piece'' of this function by $G_{M}^\text{QAM}$.

%\vspace{-0.1em}

Dashed lines in Fig.~\ref{AIRs} show the GMI of the GMI-optimized constellations for $M=16$ (blue), $M=64$ (red), and $M=256$ (green). Optimized constellations for SNRs $7$, $15$, and $21$~dB are shown as insets (bottom) in Fig.~\ref{AIRs}, where we use the name $\text{G-GS-M}$ to refer to these optimized constellations. At an AIR of $3.14$~bit/sym, {G-GS-64} gives an SNR sensitivity improvement of $0.56$~dB. This gain increases to $0.9$~dB at $5.15$~bit/sym for {G-GS-256}. See horizontal black arrows in Fig.~\ref{AIRs}.

%\vspace{-0.1em}

\begin{figure}
\centering
\includegraphics[width=1\linewidth]{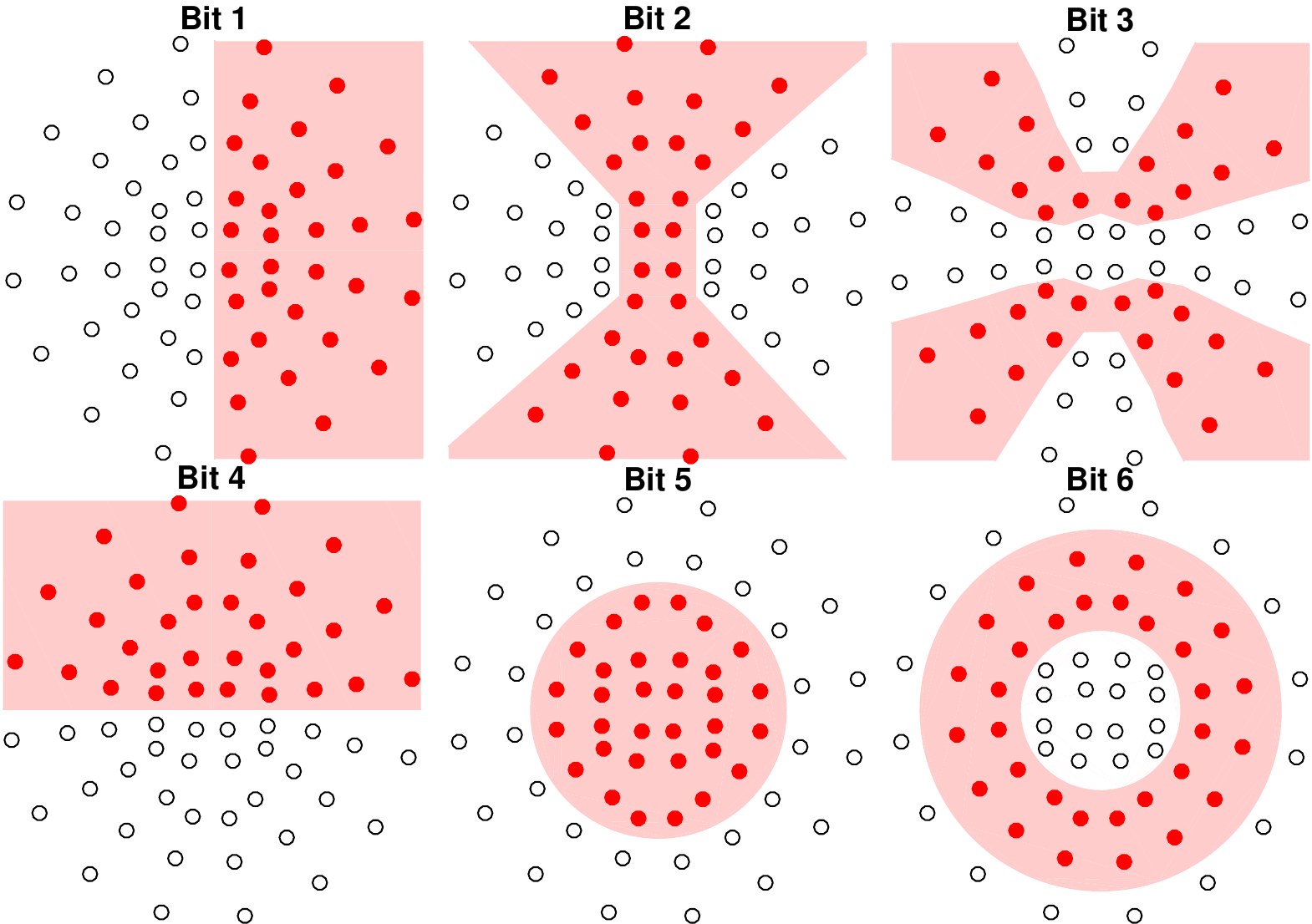}
\caption{Constellation and labeling for G-GS-64 for 15 dB. Shaded regions are the decision regions for each bit.}
\label{fig:formats_labelling}
\end{figure}

Fig. \ref{fig:formats_labelling} shows the GMI-optimized constellation {G-GS-64@15~dB} and its corresponding binary labeling. This constellation achieves $4.8$~bits/2D-symbol at $15$~dB SNR. The shaded areas in Fig. \ref{fig:formats_labelling} show the bit-wise decision regions the demapper will use to compute soft bits, where $B_i=0$ and $B_i=1$ are shown with white and red resp. This figure shows that the resulting optimized constellation and binary labeling have a highly regular structure.

%\vspace{-0.1em}

The MI of MI-optimized GS-modulation formats is also shown in Fig.~\ref{AIRs} (dotted lines). Three optimized constellations (denoted by I-GS-M) are shown in insets in Fig.~\ref{AIRs} (optimized for  $7$, $15$ and $21$~dB). These results show that AIRs can be further improved by employing MI-optimized GS constellations. For AIRs between $3$ and $6$~bit/sym, these gains are approximately $0.3$~dB over the GMI-optimized GS constellations, and up to $1.18$~dB over square QAM. These results highlight the potential gains obtained by using BICM with iterative demapping (or nonbinary FEC) in combination with GS constellations.

%\vspace{-0.45em}
\begin{figure}[h]
\centering
\includegraphics[width=1\linewidth]{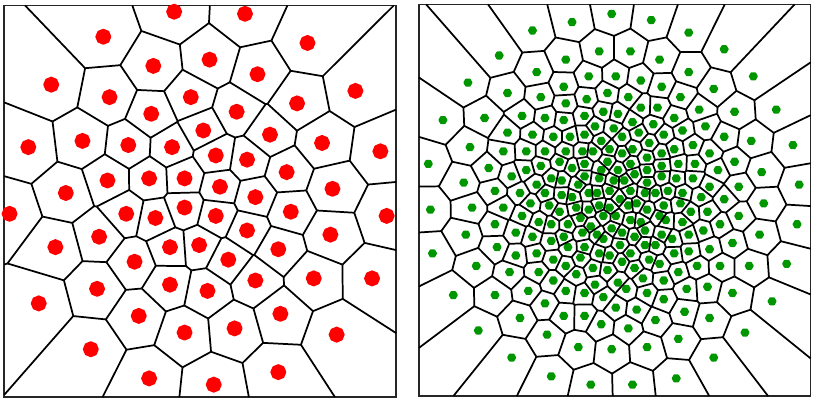}
\caption{Constellations I-GS-64 and I-GS-256 for $15$~dB.}
\label{fig:GS_MI_15dB_format}
\end{figure}

Fig. \ref{fig:GS_MI_15dB_format}  shows the  MI-optimized constellations for SNR $15$~dB with  $M=64$ and $M=256$. 
The Voronoi boundaries indicate the symbol-wise decision regions for the MI-GS formats.
For the MI-optimized constellations, these regions are shown to be more circularly symmetric than those of square QAM, which intuitively explains their higher AWGN tolerance.
%With the MI-optimized constellation, a better protection can be provided for the entire modulation formats by reducing the probability of a point added with Gaussian noise outside its Voronoi boundary.
%increases the Euclidean distance instead of the minimum Euclidean distance of constellation points  to %the symbol-wise decision region to provide better protection for all re symbol.
The optimized constellations in Fig. \ref{fig:GS_MI_15dB_format} yield MIs of $4.84$~bits/2D-symbol and $4.98$~bits/2D-symbol, resp. Therefore, the gains with respect to G-GS-64 (dashed red line in Fig.~\ref{AIRs}) constellations are $0.04$~bits/2D-symbol and $0.18$~bits/2D-symbol, resp.
\begin{figure*}[tb]
\centering
\includegraphics[width=1.01\linewidth]{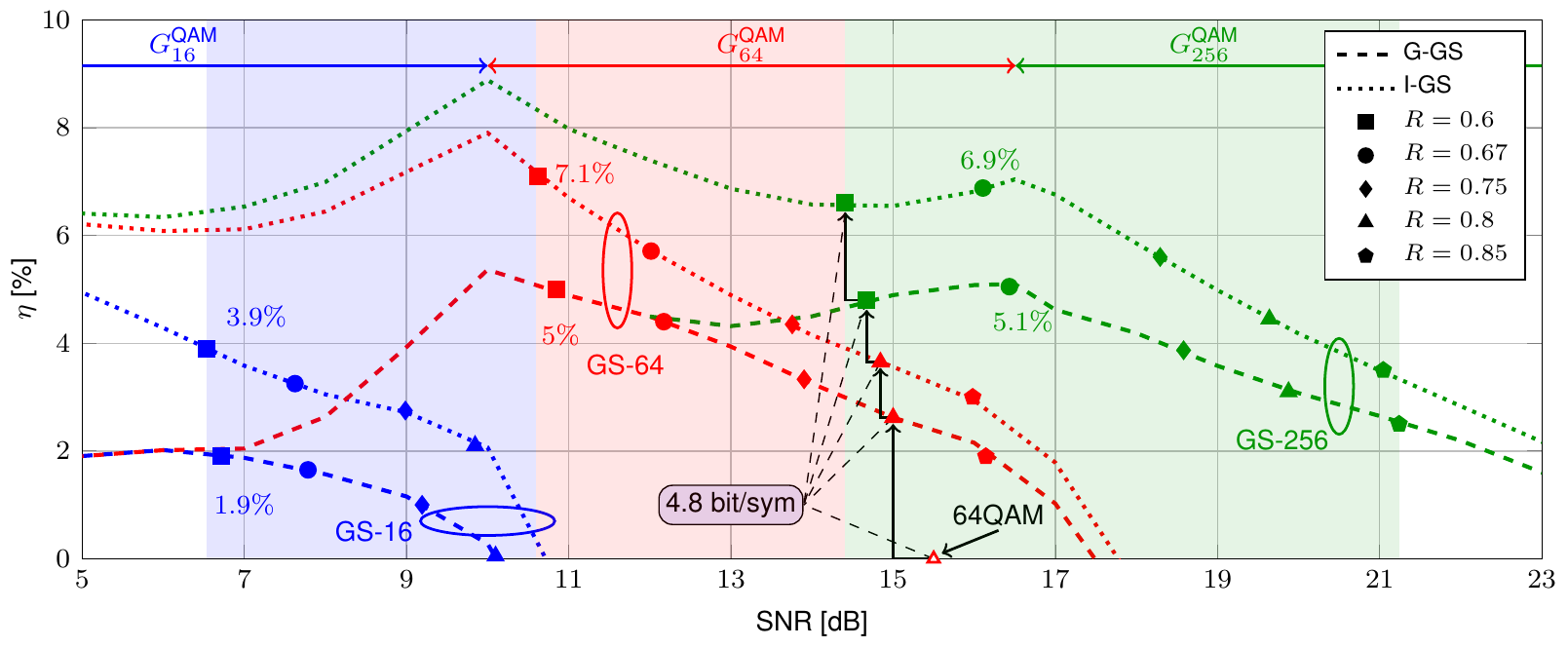}
\caption{Relative AIR  gains $\eta$ as a function of  SNR.
%  of MI/GMI-based GS-constellation with respect to the GMI of M-QAM. 
The  markers show the relative gain of GS with  FEC rates $R=[0.6 , 0.67, 0.75, 0.8, 0.85 ]$.
Color-matched shades indicate the optimum SNRs intervals for GS-16, GS-64 and GS-256.}
\label{fig:GS_Relative_Gain}
\end{figure*}

\vspace{-0.7em}
\section{Numerical Results: Relative Gains}

To better quantify the gains offered by GS constellations, here we consider \emph{relative} AIR gains with respect to
the baseline in Fig.~\ref{AIRs} (solid lines). These AIR gains are defined as $\eta= ({I^*-G^\text{QAM}})/{G^\text{QAM}}$ and $\eta= ({G^*-G^\text{QAM}})/{G^\text{QAM}}$, where $I^*$ and $G^*$ represent the AIRs obtained with the optimized constellations.

Fig.~\ref{fig:GS_Relative_Gain} shows the obtained relative AIR gains. The dashed curves show the relative gains of GMI-optimized  constellations, where the two peaks come from the chosen baseline $G^\text{QAM}$ (see horizontal arrows in Fig.~\ref{fig:GS_Relative_Gain}). We also consider the combination of practical FEC rates $0.6\leq  R\leq  0.85$ and GS modulation formats. We observe that GMI-optimized GS constellations with $M=16, 64, 256$ provide up to $1.9, 5, 5.1\%$ relative AIR gains, resp. This is shown with markers on top of dashed lines in Fig.~\ref{fig:GS_Relative_Gain}. Interestingly, these maximum gains are obtained for relatively low FEC rates. Fig.~\ref{fig:GS_Relative_Gain} also shows the relative gains of  MI-optimized GS constellations (dotted lines). The maximum gains in this case are $3.9, 7.1, 6.9\%$ and again obtained for low FEC rates.

The three shaded regions in Fig.~\ref{fig:GS_Relative_Gain} show the SNR regions that define which constellation cardinality should be chosen in order to maximize the net data rate if these particular five FEC rates were used. The optimum modulation format switches from GS-16 to GS-64 at a SNR of $10.6$~dB and from GS-64 to GS-256 at $14.4$~dB. %At SNRs higher than $16.7$~dB.

Lastly, Fig.~\ref{fig:GS_Relative_Gain} highlights five combinations of FEC and modulation formats that give the same AIR of $4.8$~bit/sym. The first one is square 64QAM (white-filled triangle), which requires $15.5$~dB SNR. By using G-GS-64 with $R=0.8$ (red triangle) and G-GS-256 with $R=0.6$ (green square), the required SNR is lowered to $15$ and $14.84$~dB, resp. The last two cases are obtained by considering MI. In this case, the required SNR is further reduced  to $14.67$ and $14.41$~dB for $R=0.8$ and $R=0.6$, resp.

% G-GS-64 with $R=0.8$ (red triangle) and G-GS-256 with $R=0.6$  (green square) are highlighted to show the same AIR of 4.8 bit/sym and compared to 64QAM with $R=0.8$ (unfilled red triangle) in Fig.~\ref{fig:GS_Relative_Gain}. We can observe that it is more spectrally efficient to use the  G-GS-64 and  I-GS-64 instead of 64QAM with the same rate 0.8 to reduce the required SNR from 15.5 dB to 15 dB and 14.84 dB, respectively.
% While  the G-GS-256  and I-GS-256 constellations with a lower code rate $R=0.6$ FEC is employed,  the required SNR can  be further reduced the  to 14.67 dB and 14.41 dB (equivalently double the relative gain).

\vspace{-0.65em}
\section{Conclusions}
We optimized the geometry of constellations based on mutual information and generalized mutual information. The reported constellations result in higher data rates for a wide range of SNRs, different constellation cardinalities, and different FEC overheads. 
To achieve larger gains, low FEC rates should be considered.
%The optimized constellations can be used as flexible formats to support different required transmission properties in flexible optical networks.

{\small
\noindent \textbf{Acknowledgements:} This work was supported by Huawei France through the NLCAP project.}

\vspace{-0.8em}

%-------------------------------------------------- Literature -------------------------------------------------------%
\balance
\bibliographystyle{abbrv}
\begin{spacing}{1.25}

\end{spacing}
\vspace{-4mm}

%%%%%%%%%%%%%%%%%%%%%%%%%%%%%%%%%%%%%%%%%%%%%
%---------------------------------------------- End of Document -----------------------------------------------%
\end{document}